\documentclass[a4paper,11pt]{article}

\usepackage{jheppub} 

\usepackage[T1]{fontenc} 

\usepackage{booktabs} 
\usepackage{array} 
\usepackage{paralist} 
\usepackage{subfig} 

\usepackage{amsfonts}

\DeclareMathOperator{\Tr}{Tr}
\DeclareMathOperator{\pslash}{\displaystyle{\not}}

\title{\boldmath Boundary Operators of BCFW Recursion Relation}

\author[a]{Qingjun Jin}
\author[a,b]{Bo Feng}

\affiliation[a]{Zhejiang Institute of Modern Physics,
Zhejiang University, 
Hangzhou, 310027, P. R. China}
\affiliation[b]{Center of
Mathematical Science, Zhejiang University, Hangzhou, China}

\emailAdd{qingjun@zju.edu.cn}
\emailAdd{b.feng@cms.zju.edu.cn}

\abstract{We show that boundary contributions of BCFW recursions can be interpreted as the form factors of some composite operators which we call 'boundary operators'. The boundary operators can be extracted from the operator product expansion of deformed fields. We also present an algorithm to compute the boundary operators using path integral.}

\begin{document} 
\maketitle
\flushbottom

\section{Introduction}
\label{section1}

It is well known that  BCFW recursion relations \cite{Britto:2004ap, Britto:2005fq} (see also reviews \cite{Bern:2007dw, Feng:2011np, Elvang:2013cua} ) fails to capture a piece of scattering amplitude, usually called boundary contribution, when the scattering amplitude does not have the desired vanishing large $z$-behavior under a given BCFW deformation. After years of development, there are several methods proposed to solve this problem, such as introducing auxiliary fields to eliminate boundary terms \cite{Benincasa:2007xk, Boels:2010mj},
analyzing Feynman diagrams to isolate boundary terms \cite{Feng:2009ei, Feng:2010ku, Feng:2011twa},
expressing boundary terms as roots of amplitudes \cite{Benincasa:2011kn, Benincasa:2011pg, Feng:2011jxa}, and
collecting factorization limits to interpolate boundary terms \cite{Zhou:2014yaa}. However, these methods are applicable only to limited types of theories, and the evaluation of boundary contributions in more general theories remains a difficult task. Recently two  ideas have been proposed to solve  boundary contributions. In \cite{Feng:2014pia,Jin:2014qya,Feng:2015qna}, multiple steps of BCFW-deformations have been used to track boundary with certain poles one by one, while in \cite{Cheung:2015cba}, multiple legs have been used to do the deformation. These two methods can be applied to  broader  theories although the pure polynomial part of boundary contributions remains a challenge.

In \cite{Jin:2014qya} it was found that with proper choices of deformations boundary contributions satisfy similar recursion relations as scattering amplitudes. Based on the observation, a systematical algorithm was proposed in which scattering amplitudes are computed with multiple steps of recursions. The first step is the usual BCFW recursion relation, which leaves out a boundary contributions\footnote{$B^{\langle 1|n]}$ means the boundary contribution to the amplitude under the $\langle 1|n]$ deformation. $B^{\langle 12|n]}$ means the boundary contribution to $B^{\langle 1|n]}$ under the $\langle 2|n]$ deformation.} $B^{\langle 1|n]}$. The second step is a recursive computation of $B^{\langle 1|n]}$, which leaves out a new boundary contribution $B^{\langle 12|n]}$, and so on. A $k$-step recursion yields the complete amplitude if $B^{\langle 1\cdots k|n]}$ vanishes. Especially it has been  proved that scattering amplitudes of 4D renormalizable theories can be computed by (at most) 4 steps of recursions.

The multiple step recursion method could be further refined in several directions. First, in \cite{Jin:2014qya} the large $z$-behavior is determined through a careful analysis of Feynman diagrams, which was already cumbersome for the examples considered therein. More convenient and systematic methods are needed for the discussion of models with more complicated matter content and interactions. The next problem is under what condition could we decide the large $z$-behavior of a deformation through some direct tests of simple amplitudes. Suppose under a certain deformation, the large $z$-behavior of any amplitude with number of legs $n<N$ is $\mathcal{O}(\frac{1}{z})$, then it is tempting to speculate that the same goes for any $n$. However, this is not always true, at least if the Lagrangian contains a $m\ge N$ point vertex which only contributes to amplitudes with $n\ge m$. Last but not least, in many theories even after exploiting all possible deformations, the last boundary contributions $B^{\langle 1\cdots n-2|n]}$ still do not vanish.
The first way to solve the problem is, of course, developing a supplementary method which computes $B^{\langle 1\cdots n-2|n]}$ directly. Another avenue to attack this problem is to explore whether $B^{\langle 1\cdots n-2|n]}$ can be discarded at some limit(for example suppressed by some large energy scale in an effective field theory) even if $B^{\langle 1\cdots n-2|n]}\neq 0$, so that the recursion terms would be a good approximation of the complete amplitude.

Moving forward in any of the directions above requires a more rigorous understanding and transparent formulation of boundary contributions. In this paper, inspired by operator product expansion (OPE), we relate boundary contributions to some composite operators which we call 'boundary operators'. Boundary contributions can be expressed as the form factors involving the corresponding boundary operators and the undeformed external states. The employment of this new formulation helps to solve all the questions above. For the first two questions, the large $z$-behavior can be read directly from the OPE.  The vanishing of boundary operator is equivalent to the vanishing of boundary contributions for all amplitude under the corresponding shift. For the last question, the last boundary contribution can be read directly from the corresponding boundary operator, since there is only one undeformed external leg left.

Boundary operators serve as a bridge connecting on shell and off shell quantities together. During the last two decades the computation of scattering amplitudes has been facilitated by the newly developed unitarity based methods, but these new methods usually do not apply to off shell quantities like form factors and correlation functions. So although not the focus of the current paper, our result will open a new window into the computation of form factors and correlation functions, and can be seen as part of the effort devoted to extract all physical quantities from scattering amplitudes.

The structure of the paper is as follows. In section \ref{section2}, we briefly review the boundary contributions, and discuss relations between boundary operators and operator product expansion(OPE). In section \ref{section3}, we present a path integral derivation of the boundary operators. In section \ref{section4}, we give some examples to demonstrate the computation of boundary operators, and in section \ref{section5}, we give some concluding remarks. In the appendixes further examples are presented.

\section{Boundary Contributions and Operator Product Expansion}
\label{section2}

A well known fact about BCFW recursion is that the large $z$-behavior of amplitudes under a given
BCFW-deformation usually only depends on the type of two deformed legs, not the type or number of undeformed legs\footnote{Of course, as pointed out in the introduction, sometimes boundary contributions of a shift  vanish accidentally for amplitudes with small $n$.}.
 For example, the large $z$-behavior in Yang-Mills theory can already be seen from 3-point or 4-point MHV gluon amplitude. This interesting 'locality' property of BCFW recursion is also inherited by multiple step recursions in \cite{Jin:2014qya}, and by $(\mathcal{I},\tilde{\mathcal{I}})$ multiple line shift discussed in \cite{Cheung:2015cba}.

This 'locality' property of large $z$-behavior is most naturally described in the language of the operator product expansion (OPE).  More explicitly, in a quantum field theory, the product of two nearby operators $\mathcal{O}_I(x)$ and $\mathcal{O}_J(y)$ can be expanded by a basis of local operators:
\begin{equation}
\mathcal{O}_I(x)\mathcal{O}_J(y)=\sum_KC_{IJ}^{\ \ \ K}(x-y)\mathcal{O}_K(y)~~~\label{ope1}
\end{equation}
In momentum space, \eqref{ope1} is translated to
\begin{equation}
\mathcal{O}_I(p_1)\mathcal{O}_J(p_2)
=\sum_KC_{IJ}^{\ \ \ K}(p_1)\mathcal{O}_K(p_1+p_2)~~~\label{ope2}
\end{equation}
where the condition that $x$ and $y$ are nearby in coordinate spaces is equivalent to the condition that $p_1$ and $p_2$ are quite split in momentum space. If we set $p_1$ to $k_1+zq$, and $p_2$ to $k_n-zq$, then the OPE holds when $z$ is large:
\begin{equation}
\mathcal{O}_I(k_1+zq)\mathcal{O}_J(k_n-zq)=\sum_KC_{IJ}^{\ \ \ K}(k_1+zq)\mathcal{O}_K(k_1+k_n)~~~~\label{ope}
\end{equation}
If the momentum of an operator is not shifted by $\pm zq$ under the BCFW shift, we will call it a soft operator. And if the momentum of an operator is shifted, we will call it a hard operator. The total momentum of two hard operators with approximately opposite momentum is soft, and \eqref{ope} tells us that the effect of these two hard fields must be described by some local soft operator. 

The r.h.s. of \eqref{ope} depends on $z$ only through some $c$-number functions $C_{IJ}^{\ \ \ K}(k_1+zq)$. Expanding  coefficients  around $z=\infty$,
\begin{equation}
C_{IJ}^{\ \ \ K}(k_1+zq)=\sum_iC_{iIJ}^{\ \ \ K}z^i\ ,~~~\label{ope-z-exp}
\end{equation}
then we have the following asymptotic expansion of \eqref{ope} at large $z$, which we call $\mathcal{G}(z)$\footnote{In the next section we will see $\mathcal{G}(z)$ can be interpreted as the Green's function of two hard fields in the background of soft fields.},
\begin{equation}
\begin{aligned}
\mathcal{G}(z)\equiv \mathcal{O}_I(k_1+zq)\mathcal{O}_J(k_n-zq)=&\sum_i \mathcal{G}_iz^i,\\
\end{aligned}\label{opez}
\end{equation}
in which $\mathcal{G}_i\equiv\sum_KC_{iIJ}^{\ \ \ K}\mathcal{O}_K(k_1+k_n)$, and each $\mathcal{G}_i$ is a local operator.

In the language of {\bf form factors}, \eqref{opez} becomes
\begin{equation}\label{opeam}
\Bigl\langle0\Bigl| \mathcal{O}_I(k_1+zq)\mathcal{O}_J(k_n-zq)\Bigr|\Phi(k_2)\cdots \Phi(k_{n-2})\Bigr\rangle
=\sum_i z^i\Bigl\langle0\Bigl| \mathcal{G}_i\Bigr|\Phi(k_2)\cdots \Phi(k_{n-2})\Bigr\rangle .
\end{equation}
If we set $\mathcal{O}_I$ and $\mathcal{O}_J$ to be two fundamental fields $\Phi(k_1+zq)$ and $\Phi(k_n-zq)$, after LSZ reduction\footnote{This means strictly speaking $\mathcal{O}_I(k_L)$ and $\mathcal{O}_J(k_R)$ are not simply the fundamental field. Instead they are fundamental fields with LSZ reduction.}, \eqref{opeam} becomes an $n$ point scattering amplitude $\mathcal{A}(\Phi(k_1),\cdots ,\Phi(k_n))$ under the $\langle 1|n]$ shift, and the boundary contribution under this $\langle 1|n]$ shift can be written as

\begin{equation}
B^{\langle 1|n]}=\langle 0|\mathcal{G}_0|\Phi(k_2)\cdots \Phi(k_{n-2})\rangle\ .~~~
\label{BO-BC}
\end{equation}

In this paper, we are mainly interested in $\mathcal{G}_0$, and for later convenience we will also denote this operator as $\mathcal{O}^{\langle 1|n]}$, and call it the {\bf boundary operator} of the BCFW shift.

Similar steps leads to the boundary operator of a multiple step shift $\mathcal{O}^{\langle 1\cdots p|n]}$,
\begin{equation}
B^{\langle 1\cdots p|n]}=\langle 0|\mathcal{O}^{\langle 1\cdots p|n]}(K_{1p})|\Phi(k_{p+1})\cdots \Phi(k_{n-2})\rangle\ .
\end{equation}
where $K_{1p}\equiv k_1+k_2+\cdots +k_p$. The only difference is instead of being a fundamental field, $\mathcal{O}_J(k_R-zq)$ in \eqref{ope} is now the boundary operator of the previous shift $\mathcal{O}^{\langle 1\cdots p-1|n]}$.

\section{Derivation of the Boundary Operators}
\label{section3}

In the last section we found that boundary operator is the $\mathcal{O}(z^0)$ order of the OPE of two deformed fields. In this section we present an algorithm to compute the boundary operators via path integral. We split each field into a hard field and a soft field, and the OPE can be interpreted as the Green's function of these two hard fields, and can be evaluated exactly at tree level.

The boundary operator of a multiple step BCFW deformation corresponds to the OPE of a fundamental field and the boundary operator of the previous deformation. In order to have a uniform description of boundary operators of the first deformation and the later deformations, In subsection \ref{section3.3}, we give a uniform description of boundary operators of the first deformation and the later deformations by introducing the boundary fields. The boundary operators of later deformations correspond to the OPE of a fundamental field and a boundary field.

\subsection{A Path Integral Representation of the Boundary Operators}
\label{section3.1}

In this subsection we present a path integral presentation of boundary operators. Consider the following $n$-point correlation function in a general quantum field theory:
\begin{equation}
G_n(z)=\langle\Phi(\hat{k}_1)\Phi(k_2)\cdots\Phi(\hat{k}_n)\rangle=\int D\Phi \exp\left(iS[\Phi]\right)\Phi(\hat{k}_1)\Phi(k_2)\cdots\Phi(\hat{k}_n).
\end{equation}

In order for the OPE \eqref{opez} to hold, the momentum of the deformed particles $\Phi(\hat{k}_1)$ and $\Phi(\hat{k}_n)$ must be very large. Following Wilson's idea, we can split the field into a high energy (hard) part\footnote{In \cite{ArkaniHamed:2008yf}, fields are split into background and a perturbation parts in order to assess boundary contributions. Eq. \eqref{split} serves as an interpretation of this splitting, and we are indebted to Mingxing Luo for pointing this out.} $\Phi^{\Lambda}$, and a low energy (soft) part (which is still denoted by $\Phi$),
\begin{equation}\label{split}
\Phi\rightarrow \Phi+\Phi^{\Lambda} \ .
\end{equation}
Under the splitting,
the action $S[\Phi]$ becomes
\begin{equation}
S[\Phi+\Phi^{\Lambda}]=S[\Phi]+S_1^{\Lambda}[\Phi^{\Lambda},\Phi]+S_2^{\Lambda}[\Phi^{\Lambda},\Phi]
+\cdots
\end{equation}
where $S_1^{\Lambda}[\Phi^{\Lambda},\Phi]$ contains terms linear in $\Phi^{\Lambda}$, and $S_2^{\Lambda}[\Phi^{\Lambda},\Phi]$ contains terms quadratic in $\Phi^{\Lambda}$, and so on.
Now we choose the energy scale $\Lambda$ to satisfy $|zq|\sim\Lambda\gg |k_i|$. Shifted lines (propagators or external legs) carry momentum $p\pm zq$, thus correspond to the hard field $\Phi^{\Lambda}$, while undeformed lines correspond to the soft field $\Phi$. 

\begin{figure}[htb]
\centering
\includegraphics[scale=0.7]{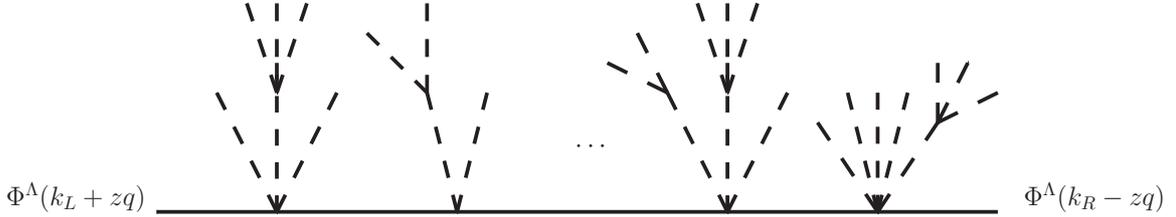}
\caption{A diagram contributing to the deformed amplitude(correlation function). Solid lines represent hard $\Phi^{\Lambda}$ fields, while dashed lines represents soft $\Phi$ field. Each vertex in the diagram contains zero or two hard fields.}
\label{fig:skeleton}
\end{figure}

A typical tree diagram containing two hard external legs and multiple soft external legs is shown in Figure  \ref{fig:skeleton}. Obviously all hard fields lie in the line connecting the shifted momenta $k_L+zq$ and $k_R-zq$, and each vertex in the diagram contains either zero or two hard fields. This means only $S[\Phi]$ and $S_2^{\Lambda}[\Phi^{\Lambda},\Phi]$ contribute to $G_n(z)$ at tree level.

Then the shifted correlation function can be evaluated as
\begin{equation}
\begin{aligned}
G_n(z)=&\int D\Phi D\Phi^{\Lambda}
 \exp\left(iS[\Phi]+iS_2^{\Lambda}[\Phi^{\Lambda},\Phi]\right)
\Phi^{\Lambda}_1\Phi^{\Lambda}_n\Phi_2\cdots \Phi_{n-1}\\
=&\int D\Phi  \exp\left(iS[\Phi]\right)
\left[\int D\Phi^{\Lambda}\exp\left(iS_2^{\Lambda}[\Phi^{\Lambda},\Phi]\right)
\Phi^{\Lambda}_1\Phi^{\Lambda}_n\right]\Phi_2\cdots \Phi_{n-1}\\
\end{aligned}\label{gz}
\end{equation}
where for simplicity we used the notation
\begin{equation}
\Phi^{\Lambda}_1\equiv\Phi^{\Lambda}(k_1+zq),\ \Phi^{\Lambda}_n\equiv\Phi^{\Lambda}(k_n-zq),\
\Phi_i\equiv\Phi(k_i)\ .
\end{equation}
In \eqref{gz}, only terms in the square brackets depend on $z$, which is the Green's function of two hard fields in the background of soft fields, and will be denoted by
\begin{equation}\label{zz}
G(z)=-i\int D\Phi^{\Lambda}\exp\left(iS_2^{\Lambda}[\Phi^{\Lambda},\Phi]\right)
\Phi^{\Lambda}_1\Phi^{\Lambda}_n
\end{equation}
and can be interpreted as the OPE of $\Phi^{\Lambda}_1$ and $\Phi^{\Lambda}_n$. After doing LSZ reduction to both deformed fields, $G(z)$ will become $\mathcal{G}(z)$ in \eqref{opez}.

\subsection{Evaluation of the Path Integral}
\label{section3.2}

In \eqref{zz}, $S_2^{\Lambda}[\Phi^{\Lambda},\Phi]$ only contain terms quadratic in $\Phi^{\Lambda}$, and $G(z)$ can be evaluated exactly.  Now we must be more careful about the components of $\Phi$. In 4-D theories, there are real fields (gauge fields, Majorana spinors and real scalars) and complex fields (Weyl fermions and complex scalars). Suppose the theory has $M$ real fields $\varphi^I$ and $N$ complex fields $\phi^A$, to simplify our notation we combine them into a single field $\Phi^{\alpha}$, and combine the hard fields into a single hard field $H^{\alpha}$,
\begin{equation}
\begin{aligned}
&\Phi^{\alpha}\equiv\begin{pmatrix}
\varphi^I \\
\phi^A \\
\bar{\phi}_A
\end{pmatrix},\
H^{\alpha}\equiv\begin{pmatrix}
\hat{\varphi}^I \\
\hat{\phi}^A \\
\hat{\bar{\phi}}_A
\end{pmatrix}.\ \\
\end{aligned}
\end{equation}
The Hermitian conjugation of $\Phi^{\alpha}$
is $\Phi^{\dagger}_{\alpha}=\begin{pmatrix}
\varphi^I &\bar{\phi}_A&\phi^A
\end{pmatrix}$,
which can be related to $\Phi^{\alpha}$ as $\Phi^{\alpha}=T^{\alpha\beta}\Phi^{\dagger}_{\beta}$  with the matrix
\begin{equation}
\begin{aligned}
&T^{\alpha\beta}=\begin{pmatrix}
I_M & 0& 0 \\
0 & 0 & I_N \\
0 & I_N & 0
\end{pmatrix}.\ \\
\end{aligned}
\end{equation}
as a metric to move indices up. With these notations, the quadratic term in the Lagrangian can be written as
\begin{equation}
\begin{aligned}
L_2^{\Lambda}=& \frac{1}{2}H^{\dagger}_{\alpha}{\cal D}^{\alpha}_{\ \ \beta}H^{\beta},~~~~~~
{\cal D}^{\alpha}_{\ \ \beta}=\frac{\delta^2}{\delta \Phi^{\dagger}_{\alpha}\delta \Phi^{\beta}}L,\\\
\end{aligned}
\end{equation}
in which $D^{\alpha}_{\ \ \beta}$ is a Hermitian operator. Following the standard procedure of computing generating functions, we add a source field $J^{\alpha}$ for $H^{\alpha}$,
\begin{equation}
\begin{aligned}
S_2^{\Lambda}[J,H,\Phi]=&\int d^Dx\Bigl(\frac{1}{2}H^{\dagger}_{\alpha}{\cal D}^{\alpha}_{\ \ \beta}H^{\beta}
+J^{\dagger}_{\alpha}H^{\alpha}\Bigr).\\
\end{aligned}\label{Gaussian-1}
\end{equation}
The source field also satisfy $J^{\alpha}=T^{\alpha\beta}J^{\dagger}_{\beta}$. It is easy to see that in \eqref{Gaussian-1}, $J^{\dagger}_{\alpha}H^{\alpha}=H^{\dagger}_{\alpha}J^{\alpha}$.

 Shifting $H^{\alpha}\rightarrow H^{\alpha}-({\cal D}^{-1})^{\alpha}_{\ \ \beta}J^{\beta}$, and integrate over the hard field, we find
\begin{equation}
\begin{aligned}
Z^{\Lambda}[J,\Phi]\equiv&\int DH\exp\left(iS_2^{\Lambda}[J,H,\Phi]\right)\\
=&Z^{\Lambda}[\Phi]\exp\left(-\frac{i}{2}\int d^Dxd^DyJ^{\dagger}_{\alpha}(x)\left({\cal D}^{-1}\right)^{\alpha}_{\ \ \beta}(x,y;\Phi)J^{\beta}(y)\right)\\
=&Z^{\Lambda}[\Phi]\exp\left(-\frac{i}{2}\int d^Dxd^DyJ^{\dagger}_{\alpha}(x)\left({\cal D}^{-1}\right)^{\alpha\beta}(x,y;\Phi)J^{\dagger}_{\beta}(y)\right),\ \\
\end{aligned}~~~~\label{general-z-1}
\end{equation}
in which $Z^{\Lambda}[\Phi]=\Bigl[\det {\cal D}(\Phi)\Bigr]^{-\frac{1}{2}}$, and in the last line we have switched the positions of the index $\beta$ using the $T$-matrix:
\begin{equation}
\left({\cal D}^{-1}\right)^{\alpha\beta}\equiv\left({\cal D}^{-1}\right)^{\alpha}_{\ \ \gamma}T^{\gamma\beta}.
\end{equation}
Notice $T_{\alpha\beta}$ can be interpreted as a metric, and we will consider ${\cal D}^{-1}$ (as well as ${\cal D}$, ${\cal D}_0$, $V$)as a tensor whose indices can be lowered or raised by $T_{\alpha\beta}$ or $T^{\alpha\beta}$.

From \eqref{zz} we have
\begin{equation}
\begin{aligned}
G^{\alpha\beta}(z)=&-i\int DH\exp\left(iS_2^{\Lambda}[H,\Phi]\right)H^{\alpha}(x)H^{\beta}(y)=i\frac{\delta^2Z^{\Lambda}[J,\Phi]}{\delta J^{\dagger}_{\alpha}(x)\delta J^{\dagger}_{\beta}(y)}\Bigr|_{J=0}\\
=&Z^{\Lambda}[\Phi]\left({\cal D}^{-1}\right)^{\alpha\beta}(x,y;\Phi)\\
\end{aligned}\label{dd1}
\end{equation}
\begin{figure}[htb]
\centering
\includegraphics[scale=0.8]{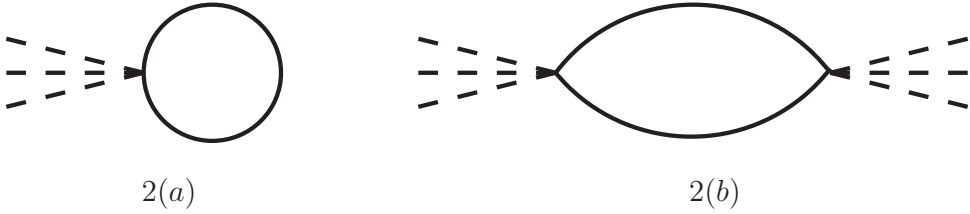}
\caption{The first two diagrams contributing to $Z^{\Lambda}[\Phi]$.}
\label{fig:sqrtd}
\end{figure}

Now we analyze these two factors in \eqref{dd1}. In general, the operator ${\cal D}(\Phi)$ can be decomposed into a free part\footnote{For example the free part $D_0$ is $\partial^2-m^2$ for a scalar field, and $i(\pslash{\partial}+m)$ for a Dirac field.} $D_0$ and an interaction part $V$,
\begin{equation}
{\cal D}^{\alpha}_{\ \ \beta}(\Phi)=(D_0)^{\alpha}_{\ \ \beta}+V^{\alpha}_{\ \ \beta}(\Phi).~~~~~\label{Dphi-decomp}
\end{equation}
The $Z^{\Lambda}[\Phi]$ term in \eqref{dd1} is a loop effect and can be set to one since in this paper we mainly discuss tree amplitude. To see this, first take the limit $\Lambda\rightarrow 0$ and $Z^{\Lambda}[\Phi]$ can be expanded as
\begin{equation}
\begin{aligned}
&Z^{\Lambda}[\Phi]\rightarrow\Bigl[\det D(\Phi)\Bigr]^{-\frac{1}{2}}\\
=&\Bigl[\det D_0\Bigr]^{-\frac{1}{2}}
\left(1-\frac{1}{2}\Tr(D_0^{-1}V)+\frac{1}{4}\Tr(D_0^{-1}VD_0^{-1}V)+\frac{1}{8}{\Tr}^2(D_0^{-1}V)+\cdots\right)\\
\end{aligned}\label{det1}
\end{equation}
The first term $\Bigl[\det D_0\Bigr]^{-\frac{1}{2}}$ in \eqref{det1} is a constant factor independent of external fields $\Phi$. The second term in the parenthesis, $-\frac{1}{2}\Tr(D_0^{-1}V)$ corresponds to Figure \ref{fig:sqrtd}(a). The third term, $\frac{1}{4}\Tr(D_0^{-1}VD_0^{-1}V)$, corresponds to Figure \ref{fig:sqrtd}(b). And the fourth term corresponds to a Feynman diagram with two copies of Figure \ref{fig:sqrtd}(a). With non-zero $\Lambda$, the path integral in $Z^{\Lambda}[\Phi]$ cannot be evaluated exactly, but the analysis above suggests
\begin{equation}
\begin{aligned}
&Z^{\Lambda}[\Phi]\sim 1-\frac{1}{2}{\Tr}_{\Lambda}(D_0^{-1}V)+\frac{1}{4}{\Tr}_{\Lambda}(D_0^{-1}VD_0^{-1}V)+\frac{1}{8}{\Tr}_{\Lambda}^2(D_0^{-1}V)+\cdots\\
\end{aligned}\label{det2}
\end{equation}
in which ${\Tr}_{\Lambda}$ means the loop integration is over the hard region $|l|>\Lambda$. In all, the $Z^{\Lambda}[\Phi]$ term is a loop effect which do not contribute in a tree level calculation, and from now on we will set it to one.

Now let us turn to the factor $\left({\cal D}^{-1}\right)^{\alpha\beta}(x,y;\Phi)$. Using the decomposition \eqref{Dphi-decomp}, the operator ${\cal D}_{\alpha\beta}$ can be formally inverted as
\begin{equation}
\begin{aligned}
{\cal D}^{-1}(\Phi) = & \left(D_0+V(\Phi)\right)^{-1}= \left(D_0[1+D_0^{-1}V(\Phi)]\right)^{-1}=
[1+D_0^{-1}V(\Phi)]^{-1} D_0^{-1}\\
= & \sum_{k=0}^{\infty} (-1)^k[D_0^{-1}V(\Phi)]^{k} D_0^{-1}
=\sum_{k=0}^{\infty} (-1)^kD_0^{-1}[V(\Phi)D_0^{-1}]^{k} \\
\end{aligned}\label{D-inverse-1}
\end{equation}

 Now we move to the OPE $\mathcal{G}(z)$, for which we need to do  LSZ reductions for fields  $H(k_1+zq)$ and $H(k_n-zq)$, i.e., we need to multiply the inverse of
 propagators and multiply corresponding external wave functions  $\epsilon_1 , \epsilon_n$. This action is given by
 the operator $\epsilon^{i}_{\alpha_i}(-iD_0)^{\alpha_i}_{ \ \ \beta_i} H^{\beta_i}$. Putting it back to \eqref{D-inverse-1} we get the expansion\footnote{We subtract a $D_0^{-1}$ from ${\cal D}^{-1}(\Phi)$ in \eqref{exp-exp-1} because after canceling one $D_0$, another $D_0$ acts on on-shell wave function $\epsilon_1,\epsilon_n$ will give zero.}
\begin{equation}
\begin{aligned}
\mathcal{G}(z)=&\epsilon^{1}_{\alpha_1}(-iD_0)^{\alpha_1}_{ \ \ \beta_1}\epsilon^{n}_{\alpha_n}(-iD_0)^{\alpha_n}_{ \ \ \beta_n}\left({\cal D}^{-1}(\Phi)-D_0^{-1}\right)^{\beta_1\beta_n}\\
=&-\epsilon^{1}_{\alpha_1}\epsilon^{n}_{\alpha_n}\Bigl[D_0\left({\cal D}^{-1}(\Phi)-D_0^{-1}\right)D_0\Bigr]^{\alpha_1\alpha_n}\\
=&\epsilon^{1}_{\alpha_1}\epsilon^{n}_{\alpha_n}\Bigl[\sum_{k=0}^{\infty} (-1)^kV[D_0^{-1}V]^{k} \Bigr]^{\alpha_1\alpha_n}\\
=&\epsilon^{1}_{\alpha_1}\epsilon^{n}_{\alpha_n}\Bigl[V^{\alpha_1\alpha_n}-V^{\alpha_1\beta_1}(D_0^{-1})_{\beta_1\beta_2}V^{\beta_2\alpha_n}+\cdots \Bigr]\\
\end{aligned}\label{exp-exp-1}
\end{equation}
From \eqref{exp-exp-1} we see that it is most convenient to derive $\mathcal{G}(z)$ from ${\cal D}$ with two upper indices,
\begin{equation}
\begin{aligned}
&D^{\alpha\beta}=(D_0)^{\alpha\beta}+V^{\alpha\beta}\\
&D^{\alpha\beta}=T^{\alpha\gamma}T^{\beta\delta}\frac{\delta^2L}{\delta \Phi^{\gamma}\delta \Phi^{\delta}}
=\frac{\delta^2L}{\delta \Phi^{\dagger}_{\alpha}\delta \Phi^{\dagger}_{\beta}}\\
\end{aligned}\label{D-up-index}
\end{equation}

It would be worthwhile to stress again that $\mathcal{G}(z)$ is the complete OPE which contains all $z$-dependence, not only the ${\cal O}(z^0)$ order. The $\mathcal{O}(z^n)$ order of $\mathcal{G}(z)$ is the operator which corresponds to the $\mathcal{O}(z^n)$ order of the deformed amplitude. For example, the $\langle g^-(k_1)|g^+(k_n)]$ deformation in Yang-Mills theory is $\mathcal{O}(z^3)$, and the $\mathcal{O}(z^3)$ order of the deformed amplitude equals to the form factor of the operator $\mathcal{G}_3$. More details on the large $z$ behavior of Yang-Mills amplitudes can be found in Appendix \ref{appendixB}.

\subsection{Boundary Operator of the Second Deformation}
\label{section3.3}

In the previous subsection we discussed the derivation of boundary operators of the first deformation $\langle1|n]$, and in this subsection we will discuss the boundary operators of the successive deformations. The boundary operator of the $k$-th shift $\langle 12\cdots k|n]$ is given by the OPE of $H_{k}$ and the boundary operator of the previous shift $\mathcal{O}^{\langle 12\cdots k-1|n]}$. For example, when $k=2$,
\begin{equation}\label{zzk+1}
G(z)=-i\int DH\exp\left(iS_2^{\Lambda}[H,\Phi]\right)
H_{2}\mathcal{O}^{\langle 1|n]}
\end{equation}

The $\mathcal{O}^{\langle 1|n]}$ in \eqref{zzk+1} is a composite operator, and in order to evaluate the path integral, we will introduce a boundary fields\footnote{The boundary field $\mathcal{F}$ is an auxiliary field which is introduced so that we could have a uniform description of the first and the successive shifts. It will not appear in the final results \eqref{bok}.} $\mathcal{F}$, which allows us to rewrite \eqref{zzk+1} as a correlation function of two fundamental fields. The boundary fields $\mathcal{F}$ couples to the (Hermitian conjugate of) boundary operator of the previous deformation,
\begin{equation}
L_2(H,\mathcal{F},\Phi)=L(\Phi)+\bar{\mathcal{F}}\mathcal{O}^{\langle1|n]}+\bar{\mathcal{O}}^{\langle1|n]}\mathcal{F}.
\end{equation}

Then $G(z)$ can be rewritten as
\begin{equation}\label{zzk+2}
G(z)=-i\int DH D\mathcal{F}\exp\left(iS_2^{\Lambda}[H,\mathcal{F},\Phi]\right)
H_{2}\mathcal{F},
\end{equation}
where $S_2^{\Lambda}[H,\mathcal{F},\Phi]$ is defined through the Lagrangian $L_2(H,\mathcal{F},\Phi)$ by repeating the steps described in the last subsection.

 However, unlike the first shift, the boundary field $\mathcal{F}$ is non-dynamical. In order to use the result of the last subsection, we will introduce a large mass $\widetilde{M}$ to the boundary field and assume it is dynamical\footnote{The constant $\kappa$ is needed to offset the mass dimension of $\mathcal{F}$. } ,
\begin{equation}
L_{\widetilde{M}}=L(\Phi)+\bar{\mathcal{F}}\mathcal{O}^{\langle1|n]}+\bar{\mathcal{O}}^{\langle1|n]}\mathcal{F}
+\kappa\bar{\mathcal{F}}(\partial^2-\widetilde{M}^2)\mathcal{F}.~~~~\label{app-21-L}
\end{equation}
and compute the OPE of $H_{k+1}$ and $\mathcal{F}$ following the algorithm of the last section, and set $\widetilde{M}\rightarrow \infty$ in the end.

In practice, since $\widetilde{M}$ is large, the $\partial^2$ term is negligible, then we can equivalently use the Lagrangian
\begin{equation}
L_{M}=L(\Phi)+\bar{\mathcal{F}}\mathcal{O}^{\langle1|n]}+\bar{\mathcal{O}}^{\langle1|n]}\mathcal{F}
-M^2\bar{\mathcal{F}}\mathcal{F},~~~~\label{app-2-L}
\end{equation}
where $M$ is defined via $M^2=\kappa  \widetilde{M}^2$.

Using \eqref{D-up-index}, we find
\begin{equation}
\begin{aligned}
\mathcal{D}^{\alpha\beta}=&\begin{pmatrix}
\mathcal{D}_{11} & \frac{\delta \mathcal{O}^{\langle1|n]}}{\delta\Phi^{\dagger}_{\alpha}} &
\frac{\delta \bar{\mathcal{O}}^{\langle1|n]}}{\delta\Phi^{\dagger}_{\alpha}} \\
\frac{\delta \mathcal{O}^{\langle1|n]}}{\delta\Phi^{\dagger}_{\beta}} & 0 & -M^2 \\
\frac{\delta \bar{\mathcal{O}}^{\langle1|n]}}{\delta\Phi^{\dagger}_{\beta}} & -M^2 & 0
\end{pmatrix}\\
\end{aligned}\label{app-2-cal-D}
\end{equation}
where $\mathcal{D}_{11}^{\alpha\beta}=D^{\alpha\beta}
+\frac{\delta^2\bar{\mathcal{O}}^{\langle1|n]}}{\delta \Phi^{\dagger}_{\alpha}\delta \Phi^{\dagger}_{\beta}}\mathcal{F}
+\bar{\mathcal{F}}\frac{\delta^2\mathcal{O}^{\langle1|n]}}{\delta \Phi^{\dagger}_{\alpha}\delta \Phi^{\dagger}_{\beta}}$ and  $D^{\alpha\beta}=\frac{\delta^2L(\Phi)}{\delta \Phi^{\dagger}_{\alpha}\delta \Phi^{\dagger}_{\beta}}
=D_0^{\alpha\beta}+V^{\alpha\beta}$. Again $\mathcal{D}$ can be split into a free part $\mathcal{D}_0$ and an interaction part $\mathcal{V}$,
\begin{equation}
\begin{aligned}
\mathcal{D}_0=&\begin{pmatrix}
D_0 &0 &0 \\
0 & 0 & -M^2 \\
0& -M^2 & 0
\end{pmatrix},\
\mathcal{V}=&\begin{pmatrix}
\mathcal{D}_{11}-D_0 & \frac{\delta \mathcal{O}^{\langle1|n]}}{\delta\Phi^{\dagger}_{\alpha}} &
\frac{\delta \bar{\mathcal{O}}^{\langle1|n]}}{\delta\Phi^{\dagger}_{\alpha}} \\
\frac{\delta \mathcal{O}^{\langle1|n]}}{\delta\Phi^{\dagger}_{\beta}} & 0 & 0 \\
\frac{\delta \bar{\mathcal{O}}^{\langle1|n]}}{\delta\Phi^{\dagger}_{\beta}} & 0 & 0
\end{pmatrix}\\
\end{aligned}
\end{equation}
Since we are only interested in the shift $\langle \Phi^{\alpha}|\mathcal{F}]$, we only need to worry about the  $(1,2)$ entry of the matrix $\mathcal{V}(1+\mathcal{D}_0^{-1}\mathcal{V})^{-1}$. Using  the large $M^2$ limit,
\begin{equation}
\begin{aligned}
\mathcal{D}_0^{-1}=&\begin{pmatrix}
D_0^{-1} &0 &0 \\
0 & 0 & 0 \\
0& 0 & 0
\end{pmatrix}+\mathcal{O}(\frac{1}{M^2})\,
\end{aligned}
\end{equation}
it is straightforward to find
\begin{equation}
\begin{aligned}
&\left[\mathcal{V}(1+\mathcal{D}_0^{-1}\mathcal{V})^{-1}\right]_{12}
=\Bigl(1+(\mathcal{D}_{11}-D_0)D_0^{-1}\Bigr)^{-1}W+\mathcal{O}(\frac{1}{M^2})\\
\end{aligned}\label{bo2}
\end{equation}
with $W^{\alpha}=\frac{\delta \mathcal{O}^{\langle1|n]}}{\delta\Phi^{\dagger}_{\alpha}}$. Eq. \eqref{bo2} can be further simplified using the fact that there is no soft $\mathcal{F}$ or $\bar{\mathcal{F}}$ fields as external fields, so we can drop terms depend on $\mathcal{F}$ or $\bar{\mathcal{F}}$ and finally reach
\begin{equation}
\begin{aligned}
&\left[\mathcal{V}(1+\mathcal{D}_0^{-1}\mathcal{V})^{-1}\right]_{12}
\rightarrow \Bigl(1+VD_0^{-1}\Bigr)^{-1}W\\
\end{aligned}\label{bo30}
\end{equation}
Last, we need to multiply this quantity by the wavefunction of $H(k_2)$,
\begin{equation}\label{bo3}
\mathcal{G}^{\langle 12|n]}(z)=\epsilon^2_{\alpha_2}\Bigl[\Bigl(1+VD_0^{-1}\Bigr)^{-1}\Bigr]^{\alpha_2}_{\ \ \beta}W^{\beta}.
\end{equation}

Similarly, the boundary operators of the $k$-th shift, $\mathcal{O}^{\langle1\cdots k|n]}$, is given by the $\mathcal{O}(z^0)$ order of $\mathcal{G}^{\langle 1\cdots k|n]}(z)$, which has the expression
\begin{equation}\label{bok}
\mathcal{G}^{\langle 1\cdots k|n]}(z)=\epsilon^k_{\alpha_k}\Bigl[\Bigl(1+VD_0^{-1}\Bigr)^{-1}\Bigr]^{\alpha_k}_{\ \ \beta}\frac{\delta \mathcal{O}^{\langle1\cdots k-1|n]}}{\delta\Phi^{\dagger}_{\beta}}.
\end{equation}

\section{Examples}
\label{section4}
Now we demonstrate the computation of boundary operators by by several examples. We will start with the non-linear sigma model, which is the low energy effective field theory of pions, focusing on the derivation of $\mathcal{D}$ operator in the presence of multiple fundamental fields. Next, we propose a different decomposition of the operator $\mathcal{D}$, which simplifies the calculation especially when $V(z)\sim \mathcal{O}(z)$. Finally we discuss the boundary operators in a theory with Yukawa and quartic scalar interactions.

\subsection{The Non-linear Sigma Model}
\label{section4.1}

The low energy effective field theory of pions is described by the following $SU(2)$ non-linear sigma,
\begin{equation}
L=-\frac{1}{2}\frac{(\partial_{\mu}\pi)^2}{(1+\frac{\pi^2}{F^2})^2},
\end{equation}
where $\pi=(\pi^1,\pi^2,\pi^3)$, and $F$ is a parameter with mass dimension $+1$.

In order to find ${\cal D}$ first we compute the variation of the Lagrangian with respect to $\pi^a$,
\begin{equation}\label{dldpi}
\frac{\delta L}{\delta \pi^a}=-\partial_{\mu}\frac{\partial^{\mu}\pi^a}{\omega^2}
+\frac{2\pi^a(\partial_{\mu}\pi)^2}{F^2\omega^3},
\end{equation}
where we have defined
\begin{equation}
\omega=1+\frac{\pi^2}{F^2}.
\end{equation}
In this subsection, we use the convention that if a $\partial_{\mu}$ is inside a parenthesis, then this $\partial_{\mu}$ only act on fields on its right inside the parenthesis. A $\partial_{\mu}$ in a square bracket acts on all fields on its right, even outside of the square bracket.

There are two more subtleties in deriving \eqref{dldpi}. First, $\frac{\delta L}{\delta \pi^a}$ should be interpreted as a column vector
\begin{equation}
\frac{\delta L}{\delta \pi^a}=
\begin{pmatrix}
\frac{\delta L}{\delta \pi^1} \\
\frac{\delta L}{\delta \pi^2} \\
\frac{\delta L}{\delta \pi^3}
\end{pmatrix}.
\end{equation}
Second, the operator $\frac{\delta}{\delta\pi^a}$ acts from the left of $L$, i.e.
\begin{equation}
\delta L=\delta \pi^a\frac{\delta L}{\delta \pi^a}
\end{equation}

The next step is compute the variation of $\frac{\delta L}{\delta \pi^a}$ with respect to $\pi^b$, but this time $\frac{\delta}{\delta\pi^b}$ acts on the right. Thus to avoid ambiguities, we will use $\frac{\overleftarrow{\delta}}{\delta\pi^b}$ to denote this variation:
\begin{equation}
\begin{aligned}
{\cal D}^{ab}=\frac{\delta L}{\delta \pi^a}\frac{\overleftarrow{\delta}}{\delta\pi^b}=&
\delta^{ab}\partial_{\mu}\frac{1}{\omega^2}\partial^{\mu}
+\delta^{ab}\frac{2(\partial_{\mu}\pi)^2}{F^2\omega^3}
-\frac{12(\partial_{\mu}\pi)^2\pi^a\pi^b}{F^4\omega^4}\\
&+\frac{4\pi^a}{F^2\omega^3}(\partial^{\mu}\pi^b)\partial_{\mu}
+\partial_{\mu}(\partial^{\mu}\pi^a)\frac{4\pi^b}{F^2\omega^3}.\\
\end{aligned}\label{dabnonlinear}
\end{equation}
Then ${\cal D}^{ab}$ is the correct form of the differential operator satisfying
\begin{equation}
L_2^{\Lambda}=\frac{1}{2}\pi^{\Lambda a}{\cal D}^{ab}\pi^{\Lambda b}.
\end{equation}

Now splitting  ${\cal D}^{ab}$ into a free part $D_0^{ab}=\partial^2\delta^{ab}$ and an interaction part $V^{ab}$,
\begin{equation}
{\cal D}^{ab}=D_0^{ab}+V^{ab},
\end{equation}
we have
\begin{equation}
\begin{aligned}
V^{ab}=&
\delta^{ab}\partial_{\mu}\frac{1-\omega^2}{\omega^2}\partial^{\mu}
+\delta^{ab}\frac{2(\partial_{\mu}\pi)^2}{F^2\omega^3}
-\frac{12(\partial_{\mu}\pi)^2\pi^a\pi^b}{F^4\omega^4}\\
&+\frac{4\pi^a}{F^2\omega^3}(\partial^{\mu}\pi^b)\partial_{\mu}
+\partial_{\mu}(\partial^{\mu}\pi^a)\frac{4\pi^b}{F^2\omega^3}.\\
\end{aligned}
\end{equation}
As in previous examples, all $z$-dependence is introduced by the differentials. Thus we decompose $V^{ab}(z)$  into two parts,
\begin{equation}
\begin{aligned}
V^{ab}(z)=&V^{ab}-izW^{ab},\\
W^{ab}=&\delta^{ab}\{\partial^-,\ \frac{1-\omega^2}{\omega^2}\}
+\frac{4\pi^a}{F^2\omega^3}(\partial^-\pi^b)
+(\partial^-\pi^a)\frac{4\pi^b}{F^2\omega^3}.\\
\end{aligned}\label{wnonlinear}
\end{equation}
where $\partial^-=q^{\mu}\partial_{\mu}$, and $\{\cdot,\cdot\}$ is the anti-commutator.

Furthermore at large $z$ the propagator is 
\begin{equation}
\left(D_0^{ab}(z)\right)^{-1}=\frac{i}{2z\partial^-}\delta^{ab}+\frac{\partial^2}{4z^2(\partial^-)^2}\delta^{ab}
+\mathcal{O}(\frac{1}{z^3})
\end{equation}
From these expression, we see that since
each terms of $\mathcal{G}(z)$ in \eqref{exp-exp-1} has $k$'s $\left(D_0^{ab}(z)\right)^{-1}$ and $(k+1)$'s $V^{ab}(z)$, $\mathcal{G}(z)$ is of order $\mathcal{O}(z)$. After some careful manipulations we get  following compact form:
\begin{equation}
\begin{aligned}
\mathcal{G}(z)=&-izW(1+\frac{1}{2\partial^-}W)^{-1}+(1+W\frac{1}{2\partial^-})^{-1}\Bigl[V+W\frac{\partial^2}{4(\partial^-)^2}W\Bigr](1+\frac{1}{2\partial^-}W)^{-1}+\mathcal{O}(\frac{1}{z})\\
\end{aligned}\label{znonlinear}
\end{equation}
where the second term is the boundary operator ${\cal O}^{\langle 1|n]}$ we are looking for.

\subsection{An Alternative Decomposition of ${\cal D}$}
\label{section4.1}

In the last subsection we see that when $V(z)\sim \mathcal{O}(z)$, the boundary operator receive contribution from diagrams with arbitrary number of hard propagators, which makes the derivation of boundary operator quite involved.  In order to simplify the computation in this subsection we introduce a different decomposition,
\begin{equation}
{\cal D}={\cal D}_z+V_0,\ {\cal D}_z={\cal D}_0+V_z.
\end{equation}
in which we have absorbed $\mathcal{O}(z)$ terms of $V$ to a new differential operator ${\cal D}_z$, and $V_0\sim \mathcal{O}(z^0)$.

 Then $\mathcal{G}(z)$ can be written as\footnote{For simplicity in this subsection we will neglect wavefunctions $\epsilon^i_{\alpha_i}$.}
\begin{equation}
\begin{aligned}
\mathcal{G}(z)=&-\mathcal{D}_0(\mathcal{D}^{-1}-\mathcal{D}_0^{-1})\mathcal{D}_0\\
=&-\mathcal{D}_0\Bigl[\mathcal{D}_z^{-1}-\mathcal{D}_0^{-1}\Bigr]\mathcal{D}_0
+\mathcal{D}_0\mathcal{D}_z^{-1}V_0
(1+\mathcal{D}_z^{-1}V_0)^{-1}\mathcal{D}_z^{-1}\mathcal{D}_0\ .\\
\end{aligned}\label{zzz1}
\end{equation}
We will assume $V_z(z)\sim \mathcal{O}(z)$, then $\mathcal{D}_z^{-1}\sim\mathcal{O}(\frac{1}{z})$. In the second term of \eqref{zzz1}, $\mathcal{D}_0\mathcal{D}_z^{-1}\sim\mathcal{O}(z^0)$, and $\mathcal{D}_z^{-1}V_0\sim\mathcal{O}(\frac{1}{z})$. So in $(1+\mathcal{D}_z^{-1}V_0)^{-1}$ only the leading term contributes to the boundary operator,
\begin{equation}
\begin{aligned}
\mathcal{G}(z)\sim&-\mathcal{D}_0(\mathcal{D}_z^{-1}-\mathcal{D}_0^{-1})\mathcal{D}_0
+\mathcal{D}_0\mathcal{D}_z^{-1}V_0\mathcal{D}_z^{-1}\mathcal{D}_0\\
=&V_z(1+D_0^{-1}V_z)^{-1}+(1+V_zD_0^{-1})^{-1}V_0(1+D_0^{-1}V_z)^{-1}\\
\end{aligned}
\end{equation}
To simplify this expression further, let us define
\begin{equation}
V_z(z)=V_z+zX,\ D_0^{-1}=\frac{d_1}{z}+\frac{d_2}{z^2}+\cdots
\end{equation}
then
\begin{equation}
\begin{aligned}
&(1+D_0^{-1}V_z)^{-1}\rightarrow \Bigl(1+d_1X+\frac{d_2X+d_1V_z}{z}\Bigr)^{-1}\\
=&(1+d_1X)^{-1}-\frac{1}{z}(1+d_1X)^{-1}(d_2X+d_1V_z)(1+d_1X)^{-1}+\mathcal{O}(z^{-2})\\
\end{aligned}
\end{equation}
After a bit of calculation we find $\mathcal{G}(z)$ can be written as
\begin{equation}
\begin{aligned}
\mathcal{G}(z)=&zX(1+d_1X)^{-1}+(1+Xd_1)^{-1}\Bigl[V-Xd_2X\Bigr](1+d_1X)^{-1}+\mathcal{O}(\frac{1}{z})\\
\end{aligned}\label{zz423}
\end{equation}

Now we go back to nonlinear sigma model. We will choose
\begin{equation}
\begin{aligned}
V_z^{ab}=&
\delta^{ab}\partial_{\mu}\frac{1-\omega^2}{\omega^2}\partial^{\mu}+\frac{4\pi^a}{F^2\omega^3}(\partial^{\mu}\pi^b)\partial_{\mu}
+\partial_{\mu}(\partial^{\mu}\pi^a)\frac{4\pi^b}{F^2\omega^3},\\
V_0^{ab}=&\delta^{ab}\frac{2(\partial_{\mu}\pi)^2}{F^2\omega^3}
-\frac{12(\partial_{\mu}\pi)^2\pi^a\pi^b}{F^4\omega^4},\\
\end{aligned}
\end{equation}
and we have
\begin{equation}
\begin{aligned}
X^{ab}=&-iW^{ab},\ d_1=\frac{i}{2\partial^-},\ d_2=\frac{\partial^2}{4(\partial^-)^2}.\\
\end{aligned}\label{xd12}
\end{equation}
Plug \eqref{xd12} into \eqref{zz423}, we find the same expression as \eqref{znonlinear}.

\subsection{A Theory with Yukawa and Quartic Scalar Coupling}
\label{subsection4.3}

In this section we consider a massless theory with Yukawa and $\phi^4$ interactions:
\begin{equation}
L=\frac{1}{2}\phi\partial^2\phi+i\bar{\psi}\bar{\sigma}^{\mu}\partial_{\mu}\psi
+\frac{1}{2}\lambda\phi\psi\psi+\frac{1}{2}\bar{\lambda}\phi\bar{\psi}\bar{\psi}+\frac{g}{4!}\phi^4.
\end{equation}
For compactness, here we will focus on the first deformation, and the discussion of the boundary operators of multiple step deformations can be found in Appendix \ref{appendixA}.

The chiral fermion carries $SU(2)$ indices, and we will use the following angle and bracket notations to keep track of these indices:
\begin{equation}
|\psi\rangle=\psi_a,\ \langle\psi|=\psi^a,\ |\bar{\psi}]=\bar{\psi}^{\dot{a}},\ [\bar{\psi}|=\bar{\psi}_{\dot{a}}
\end{equation}
To find the boundary operator after the first deformation, hard fields are combined into
\begin{equation}
\Phi^{\alpha}=\begin{pmatrix}
\phi\\
|\psi\rangle \\
|\bar{\psi}]
\end{pmatrix},\
\Phi_{\alpha}^{\dagger}=\begin{pmatrix}
\phi&[\bar{\psi}| & \langle\psi|
\end{pmatrix},\
\end{equation}
and  the corresponding  operator ${\cal D}^{\alpha\beta}$ can be decomposed to following free part and interaction
part:
\begin{equation}
D_0^{\alpha\beta}=\begin{pmatrix}
\partial^2&0&0\\
0&0& i\bar{\sigma}^{\mu}\partial_{\mu}  \\
 0&-i\sigma^{\mu}\partial_{\mu}&0
\end{pmatrix},\
V^{\alpha\beta}=\begin{pmatrix}
\frac{g}{2}\phi^2
&\bar{\lambda}[\bar{\psi}|
&\lambda\langle\psi|\\
\bar{\lambda}|\bar{\psi}]&\bar{\lambda}\phi&0 \\
 \lambda|\psi\rangle&0&\lambda\phi
\end{pmatrix}.\
\end{equation}
The inverse of free part $D_0^{\alpha\beta}$ is
\begin{equation}
(D_0^{-1})_{\alpha\beta}=\begin{pmatrix}
1&0&0\\
 0&0&i\bar{\sigma}^{\mu}\partial_{\mu} \\
0&  -i\sigma^{\mu}\partial_{\mu}&0
\end{pmatrix}(\partial^2)^{-1}.
\end{equation}
To analyze the $z$-dependence, one can simply do  the replacement $\partial_{\mu}\rightarrow \partial_{\mu}-izq_{\mu}$. It is not hard to see that $V^{\alpha\beta}$ is not affected by the shift, while $(D_0^{-1})_{\alpha\beta}$ becomes
\begin{equation}
(D_0^{-1})_{\alpha\beta}(z)=\frac{1}{2}\begin{pmatrix}
\frac{1}{iz\partial^-}&0&0\\
0&0& \frac{iq\cdot \bar{\sigma}}{\partial^-}+\frac{1}{z}(\frac{\partial^2q\cdot \bar{\sigma}}{2(\partial^-)^2}-\frac{\bar{\sigma}^{\mu}\partial_{\mu}}{\partial^-})\\
0&-\frac{iq\cdot \sigma}{\partial^-}-\frac{1}{z}(\frac{\partial^2q\cdot \sigma}{2(\partial^-)^2}-\frac{\sigma^{\mu}\partial_{\mu}}{\partial^-}) &0
\end{pmatrix}+\mathcal{O}(\frac{1}{z^2}).
\end{equation}

The theory has 3 physical states, therefore there are in all 9 possible ways of BCFW shifts. Among these shifts, $\langle \psi|\bar{\psi}]$ shift is $\mathcal{O}(\frac{1}{z})$, and the corresponding boundary operator vanishes. There are 5 $\mathcal{O}(1)$ shifts, and the corresponding boundary operators are
\begin{equation}
\begin{aligned}
\mathcal{O}^{\langle \phi|\bar{\psi}]}=&\lambda\langle\psi n\rangle,\
\mathcal{O}^{\langle \psi|\phi]}=\bar{\lambda}[1\bar{\psi}]\\
\mathcal{O}^{\langle \psi|\psi]}=&\bar{\lambda}[1n]\phi-\frac{i}{2}\bar{\lambda}^2[1\bar{\psi}]\frac{1}{\partial^-}[\bar{\psi}1]\\
\mathcal{O}^{\langle \bar{\psi}|\bar{\psi}]}=&\lambda\phi\langle1n\rangle+\frac{i}{2}\lambda^2\langle n\psi \rangle\frac{1}{\partial^-}\langle n\psi \rangle\\
\mathcal{O}^{\langle \phi|\phi]}=&\frac{g}{2}\phi^2+\frac{i}{2}|\lambda|^2\left[-[\bar{\psi}|\frac{q\cdot\bar{\sigma}}{ \partial^-}|\psi\rangle+\langle\psi|\frac{q\cdot\sigma}{ \partial^-}|\bar{\psi}]\right]\\
\end{aligned}
\end{equation}
using our general formula \eqref{exp-exp-1}.

The other 3 shifts are $\mathcal{O}(z)$ and the corresponding $\mathcal{G}(z)$ operators are as follows. For $\langle \phi|\psi]$ shift we get
\begin{equation}
\begin{aligned}
\mathcal{G}(z)=&z\bar{\lambda}[1\bar{\psi}]
-\bar{\lambda}[n\bar{\psi}]+\frac{i}{4}\bar{\lambda}g\phi^2\frac{1}{\partial^-}[1\bar{\psi}]
+\frac{|\lambda|^2}{2}\langle\psi|\frac{\sigma\cdot\partial}{\partial^-}|1]\phi
+\frac{i|\lambda|^2}{2}[1n]\langle\psi n\rangle|\frac{1}{\partial^-}\phi\\
+&\frac{\bar{\lambda}^2\lambda}{4}\left(
-[\bar{\psi}|\frac{q\cdot \bar{\sigma}}{\partial^-}|\psi\rangle\frac{1}{\partial^-}[\bar{\psi}1]
+\langle\psi|\frac{q\cdot \sigma}{\partial^-}|\bar{\psi}]\frac{1}{\partial^-}[\bar{\psi}1]
-2i[\bar{\psi}1]\frac{1}{\partial^-}\phi^2
\right)+\mathcal{O}(\frac{1}{z}).\\
\end{aligned}
\end{equation}
For $\langle \bar{\psi}|\phi]$ shift we get
\begin{equation}
\begin{aligned}
\mathcal{G}(z)=&-z\lambda\langle n\psi \rangle+\lambda\langle 1\psi \rangle
-\frac{ig\lambda}{4}\langle n\psi\rangle\frac{1}{\partial^-}\phi^2
+\frac{|\lambda|^2}{2}\phi\left(i\langle1n\rangle\frac{1}{\partial^-}[1\bar{\psi}]
+\langle n|\frac{\sigma\cdot\partial}{\partial^-}|\bar{\psi}]\right)\\
+&\frac{\bar{\lambda}
  \lambda^2}{4}\left(-\langle n\psi\rangle
   \frac{1}{\partial^-} [\bar{\psi}|
    \frac{q\cdot\bar{\sigma}}{\partial^-}
   |\psi\rangle +
  \langle n\psi\rangle
   \frac{1}{\partial^-} \langle\psi|
   \frac{q\cdot\sigma }{\partial^-} |\bar{\psi}] -
2 i  \phi^2
   \frac{1 }{\partial^-} \langle n\psi\rangle\right)+\mathcal{O}(\frac{1}{z})\\
\end{aligned}
\end{equation}
Finally, for $\langle \bar{\psi}|\psi]$ shift we get
\begin{equation}
\begin{aligned}
&\mathcal{G}(z)=\frac{z|\lambda|^2}{2}\left( i \langle n\psi\rangle \frac{1}{\partial^-}
  [\bar{\psi}1] -
  \langle n| \phi  \frac{\sigma\cdot\partial}{\partial^-} \phi |1]\right)\\
&+|\lambda|^2\left(-i \langle\psi|\frac{(k_1+k_n)\cdot \sigma}{2\partial^-} |\bar{\psi}] +
 \phi    \frac{(k_1+k_n)\cdot\partial-ik_1\cdot k_n}{\partial^-} \phi \right)\\
&+\frac{|\lambda|^2}{4}\Bigl(-\frac{1}{2}g \langle n\psi\rangle \frac{1}{\partial^-} \phi^2
   \frac{1}{\partial^-} [\bar{\psi}1]+
 \bar{\lambda} \phi \langle1n\rangle \frac{1}{\partial^-} [1\bar{\psi}]
   \frac{1}{\partial^-} [\bar{\psi}1] \\
&-
  i \bar{\lambda} \phi \langle n|  \frac{ \sigma\cdot\partial}{\partial^-}
   |\bar{\psi}] \frac{1}{\partial^-} [\bar{\psi}1]+
  \lambda[1n] \langle n\psi\rangle
   \frac{1}{\partial^-} \langle\psi n\rangle
   \frac{1}{\partial^-} \phi -
 i
  \lambda \langle n\psi\rangle
   \frac{1}{\partial^-} \langle\psi|
    \frac{\sigma\cdot\partial}{\partial^-} |1]\phi \Bigr)\\
&+\frac{|\lambda|^4}{8}\Bigl(
i \langle n\psi\rangle  \frac{1}{\partial^-}
\left([\bar{\psi}| \frac{ q\cdot\bar{\sigma}}{\partial^-} |\psi\rangle
-\langle\psi| \frac{q\cdot\sigma }{\partial^-} |\bar{\psi}] \right)
\frac{1}{\partial^-} [\bar{\psi}1] -
2 \phi^2
   \frac{1}{\partial^-} \langle n\psi\rangle
   \frac{1}{\partial^-} [\bar{\psi}1] \\
&-  2\langle n\psi\rangle
   \frac{1}{\partial^-} [\bar{\psi}1]
   \frac{1 }{\partial^-} \phi^2-
  4i \phi^2 \frac{1 }{\partial^-} \phi^2\Bigr)+\mathcal{O}(\frac{1}{z})
\\
\end{aligned}
\end{equation}

The boundary operators of these 3 shifts are $\mathcal{O}(1)$ order of the corresponding $\mathcal{G}(z)$ operator and they have more complicated expressions. However, in practice we can always choose shifts which have better large $z$ behaviors and the corresponding boundary operators are much simpler.

\section{Conclusion and Discussions}
\label{section5}

To summarize, we have introduced the boundary operator as a tool to study boundary contribution of BCFW recursion relations, and presented an algorithm of deriving boundary operators. To demonstrate our algorithm, several examples have been presented to show different aspects of the algorithm, like the presentation and decomposition of operator $\mathcal{D}$, and the analysis of $z$-dependence.

Another generalization of BCFW has been discussed in a recent work \cite{Cheung:2015cba}. But instead of deforming external legs one by one,  they choose to deform multiple legs at the same time. Surprisingly, these two methods have a lot of similar features. For example, any amplitude in 4-$D$ renormalizable theories can be obtained by at most 5-line(or 4 step in our method) deformation, while amplitudes in non-renormalizable effective field theories usually do not have the desired large $z$-behavior. In the language of OPE, the multiple line shift corresponds to the products of more than two operators. The evaluation of boundary operators in this case can be much more difficult because the hard field action is no longer Gaussian. But still it is worthwhile to check whether it can be accomplished at least for simple cases like 3-line shift.

As has been mentioned in the introduction, the boundary operator method can be used to extract form factors and correlation functions from scattering amplitudes. In fact, \eqref{opez} tells us each $z$ order of the shifted amplitudes corresponds to the form factor of a local operator $\mathcal{G}_i$. 

Last, let us point out that one of the advantages of the new description of boundary contributions is, since OPE can be defined at loop level or even non-perturbatively,  there should be no obstruction to generalize boundary operators to loop level or non-perturbative settings. It would be interesting to study the loop corrections to boundary operators in supersymmetric gauge theories, where BCFW recursion relations for planar amplitudes are known even at loop level \cite{CaronHuot:2010zt, ArkaniHamed:2010kv}.

\section*{Acknowledgement}

We thank Rijun Huang and Mingxing Luo for helpful discussions and Rijun Huang for comments on the draft. This work is supported by
Qiu-Shi funding and Chinese NSF funding under contracts No.11031005,
No.11135006, No.11125523 and No.10875103, and National Basic
Research Program of China (2010CB833000).

\appendix

\section{Boundary Contributions of Multiple Step Deformations}
\label{appendixA}

In subsection \ref{subsection4.3} we discussed the boundary operators in a theory with Yukawa and quartic scalar interactions, and found only one $\mathcal{O}(\frac{1}{z})$ shift, $\langle\psi|\bar{\psi}]$. In order to compute scattering amplitudes with arbitrary configurations of external legs, in this appendix we study the
boundary operator of further deformations, so we can find a simplest way to calculate the whole amplitude.

Suppose the amplitude has at least 4 $\psi$ legs, we can use a series of $\langle\psi\cdots \psi|\psi]$ shifts to find the whole amplitude. From last section we find
\begin{equation}
\begin{aligned}
&\mathcal{O}^{\langle\psi|\psi]}=\bar{\lambda}[1n]\phi-\frac{i}{2}\bar{\lambda}^2[1\bar{\psi}]\frac{1}{\partial^-}[\bar{\psi}1]\\
\end{aligned}
\end{equation}

Following the steps in subsection \ref{section3.3}, we find
\begin{equation}
\begin{aligned}
W^{\alpha}=
\begin{pmatrix}
\frac{\delta }{\delta\phi}\mathcal{O}^{\langle\psi|\psi]}\\
\frac{\delta }{\delta\bar{\psi}_{\dot{a}}}\mathcal{O}^{\langle\psi|\psi]}\\
\frac{\delta }{\delta\psi^a}\mathcal{O}^{\langle\psi|\psi]}
\end{pmatrix}
=\begin{pmatrix}
\bar{\lambda}[1n]\\
\frac{i\bar{\lambda}^2}{2}|1]\left[\frac{1}{\partial^-}[\bar{\psi}1]+\left(\frac{1}{\partial^-}[\bar{\psi}1]\right)\right]\\
0
\end{pmatrix}\\
\end{aligned}
\end{equation}
thus we get following boundary operator
\begin{equation}
\begin{aligned}
\mathcal{O}^{\langle\psi\psi|\psi|}=&i\bar{\lambda}^2[12]\frac{1}{\partial^-}[\bar{\psi}1]\ .\\
\end{aligned}\label{offf}
\end{equation}
after the second deformation, and
\begin{equation}
\mathcal{O}^{\langle\psi\psi\psi|\psi|}=i\bar{\lambda}^2[12][13]\frac{1}{\partial^-}.\
\end{equation}
after the third deformation. Since
this operator contains no soft field,  it will not contribute to a $n$ point amplitude if $n>4$.\footnote{In general, if the boundary operator of a k-step shift contains no soft fields, the boundary contribution vanishes if $n>k$. The boundary contribution does not vanish for the case $n=k$, unless the boundary operator vanishes.} Therefore $n$-point amplitudes  with at least 4 $\psi$ legs and $n\geq 5$ (or equivalently, 4 $\bar{\psi}$ legs) can be computed by three recursion steps.

If the amplitude contains at least 4 scalars, we can use a series of $\langle\phi\cdots \phi|\phi]$ shifts. After the second deformation $\langle\phi\phi|\phi]$ shift, we find boundary operator
\begin{equation}
\begin{aligned}
W^{\alpha}=&\begin{pmatrix}
g\phi\\
-\frac{i}{2}|\lambda|^2\left[\frac{q\cdot\bar{\sigma}}{ \partial^-}|\psi\rangle+\left(\frac{q\cdot\bar{\sigma}}{ \partial^-}|\psi\rangle\right)\right]\\
\frac{i}{2}|\lambda|^2\left[\frac{q\cdot\sigma}{ \partial^-}|\bar{\psi}]+\left(\frac{q\cdot\sigma}{ \partial^-}|\bar{\psi}]\right)\right]
\end{pmatrix},\
\mathcal{O}^{\langle \phi\phi|\phi]}(z)=g\phi\\
\end{aligned}
\end{equation}
After the third deformation we   find $\mathcal{O}^{\langle \phi\phi\phi|\phi]}=g$, therefore the amplitude can be computed using three recursion steps  if $n>4$ with at least four scalars. This is consistent with the result of the scalar theory: a pure scalar amplitude would not involve any Yukawa interaction, so the amplitude should be identical with amplitudes in a $\phi^4$ theory.

If the amplitude has at least one $\bar{\psi}$ and two $\phi$ legs, with a bit calculation it can be found that the complete amplitude can be computed using a two step recursion $\langle\phi\phi|\bar{\psi}]$.

In summary, any amplitude with $n>4$ in this theory can be computed using one of these 4 types of shifts: $\langle \psi|\bar{\psi}]$, $\langle \psi\psi\psi|\psi]$($\langle \bar{\psi}|\bar{\psi}\bar{\psi}\bar{\psi}]$), $\langle\phi\phi|\bar{\psi}]$($\langle\psi|\phi\phi]$) and $\langle \phi\phi\phi|\phi]$. Suppose the amplitude has least a $\psi\bar{\psi}$ pair, then $\langle \psi|\bar{\psi}]$ is a good one step shift. Pure bosonic amplitudes can be computed using a $\langle \phi\phi\phi|\phi]$ shift. For other types of amplitude, without loss of generality, we can assume there are no $\bar{\psi}$ legs. If the amplitude  contains more than 4 $\psi$ legs, it can be computed using a $\langle \psi\psi\psi|\psi]$ shift. If the amplitude only has 2 $\psi$ legs, since $n>4$, there are more than 2 $\phi$ legs, and the amplitude can be computed with a $\langle\psi|\phi\phi]$ shift.

\section{Yang-Mills Theory in $D$ Dimensions}
\label{appendixB}

In \cite{ArkaniHamed:2008yf,Cheung:2008dn}, the boundary contribution in Yang-Mills theory is discussed. The analysis of large $z$ behavior relies on a "spin Lorentz symmetry" of the Lagrangian aftering imposing a special gauge. With our new method, the large $z$ behavior of the deformed amplitude can be read directly from the OPE \eqref{opeam}, which can be computed explicitly. 

Let us consider the Yang-Mills theory in spacetime dimension $D>4$:
\begin{equation}
L=-\frac{1}{4}F_{\mu\nu}^aF^{\mu\nu a}\ .
\end{equation}
Splitting Yang-Mills field $A_{\mu}$ into
\begin{equation}
A_{\mu}\rightarrow A_{\mu}+A^{\Lambda}_{\mu}
\end{equation}
and adding the gauge-fixing term\footnote{Most conventions in this subsection follow  \cite{ArkaniHamed:2008yf} and \cite{Cheung:2008dn}.},
\begin{equation}\label{bggf}
L_{gf}=-\frac{1}{2}(\mathcal{D}_{\mu}A^{\Lambda \mu a})^2,
\end{equation}
we find the following expression for $L_2^{\Lambda}$:
\begin{equation}
\begin{aligned}
L_2^{\Lambda}
=&-\frac{1}{2}\mathcal{D}_{\mu}A^{\Lambda a}_{\nu}\mathcal{D}^{\mu}A^{\Lambda\nu a}
-gf^{abc}A^{\Lambda a}_{\mu}A^{\Lambda b}_{\nu}F^{\mu\nu c}\\
=&\frac{1}{2}A^{\Lambda a}_{\mu}\left(\eta^{\mu\nu}(\mathcal{D}^2)^{ab}-2gf^{abc}F^{\mu\nu c}\right)A^{\Lambda b}_{\nu}
\\
\end{aligned}\label{ordera2}
\end{equation}
The operator ${\cal D}$ in this case is dressed with both spacetime and $SU(N)$ indices,
\begin{equation}
\begin{aligned}
D^{\mu a;\nu b}=&\eta^{\mu\nu}\delta^{ab}\partial^2-V^{\mu a;\nu b},\\
V^{\mu a;\nu b}=&-gf^{abc}\Bigl(2F^{\mu\nu c}+\eta^{\mu\nu}\{\partial_{\alpha},A^{\alpha c}\}\Bigr)
-g^2f^{ace}f^{bde}A_{\alpha}^cA^{\alpha d}\eta^{\mu\nu}.\\
\end{aligned}
\end{equation}
Denoting the inverse of the ${\cal D}$ operator as ${\cal D}^{-1}_{\mu a;\nu b}$,  after the LSZ reduction we find
\begin{equation}
\begin{aligned}
&\partial^2\left({\cal D}^{-1}_{\mu a;\nu b}-\eta_{\mu\nu}\delta_{ab}(\partial^2)^{-1}\right)\partial^2
=V_{\mu a;\nu b}-V_{\mu a;\alpha c}(\partial^2)^{-1}V^{\alpha c}_{\ \ \ ;\nu b}+\cdots\\
\end{aligned}
\end{equation}
Now we consider the large $z$ behavior.
Under the shift $\langle1|n]$, we have
\begin{equation}
\begin{aligned}
&V^{\mu a;\nu b}(z)= V^{\mu a;\nu b}+ 2izgf^{abc}\eta^{\mu\nu}q_{\alpha}A^{\alpha c},\\
\end{aligned}
\end{equation}
i.e., $V^{\mu a;\nu b}(z)$ is of order $\mathcal{O}(z)$. However, as discussed in \cite{ArkaniHamed:2008yf} the $\mathcal{O}(z)$ piece vanishes when light cone gauge are imposed on the soft fields\footnote{We are imposing background gauge to high energy modes, and light cone gauge to low energy modes of the gauge field.  Around the energy scale $\Lambda$, a smooth function can be used to link two different gauge fixing terms.}. Most terms in \eqref{exp-exp-1} are actually at most $\mathcal{O}(\frac{1}{z})$, except the leading term, where light cone gauge can not be imposed on the single $A$ field. We will denote the leading term by $\mathcal{G}_1(z)$. Following the conventions for polarization vectors in \cite{ArkaniHamed:2008yf} we find
\begin{equation}
\begin{aligned}
&\mathcal{G}_1(z)=
\begin{pmatrix}
 \epsilon^-_{\mu}(\hat{k}_1)\\
\epsilon^+_{\mu}(\hat{k}_1) \\
\epsilon^i_{\mu}(\hat{k}_1)
\end{pmatrix}
V^{\mu a;\nu b}(z)
\begin{pmatrix}
\epsilon^+_{\nu}(\hat{k}_n) &
 \epsilon^-_{\nu}(\hat{k}_n)&
 \epsilon^j_{\nu}(\hat{k}_n)
\end{pmatrix}\\
=&\begin{pmatrix}
\bar{q}_{\mu} +zk_{n\mu}\\
q_{\mu}\\
\epsilon^i_{\mu}(\hat{k}_1)
\end{pmatrix}
\left(V^{\mu a;\nu b}+ 2izgf^{abc}\eta^{\mu\nu}q_{\alpha}A^{\alpha c}\right)
\begin{pmatrix}
\bar{q}_{\nu}-zk_{1\nu}&q_{\nu}&
 \epsilon^j_{\nu}(\hat{k}_n)
\end{pmatrix}\\
=&\begin{pmatrix}
\bar{q}_{\mu} +zk_{n\mu}\\
-\frac{1}{z}k_{1\mu}\\
\epsilon^i_{\mu}(k_1)
\end{pmatrix}
\left(V^{\mu a;\nu b}+ 2izgf^{abc}\eta^{\mu\nu}q_{\alpha}A^{\alpha c}\right)
\begin{pmatrix}
\bar{q}_{\nu}-zk_{1\nu}&\frac{1}{z}k_{n\nu}&
 \epsilon^j_{\nu}(k_n)
\end{pmatrix}\\
=&
\begin{pmatrix}
\mathcal{G}_1^{-+}(z)&0
&(\bar{q}_{\mu} +zk_{n\mu})V^{\mu a;\nu b}\epsilon^j_{\nu}\\
0&0&0\\
\epsilon^i_{\mu}V^{\mu a;\nu b}(\bar{q}_{\nu}-zk_{1\nu})
&0
&\epsilon^i_{\mu}(V^{\mu a;\nu b}+ 2izgf^{abc}\eta^{\mu\nu}q_{\alpha}A^{\alpha c})\epsilon^j_{\nu}\\
\end{pmatrix}+\mathcal{O}(\frac{1}{z})\\
\end{aligned}\label{m1}
\end{equation}
where we have chosen $\epsilon^i_{\mu}$ as normalized constant vectors satisfying
\begin{equation}
\eta^{\mu\nu}\epsilon^i_{\mu}\epsilon^j_{\nu}=\delta^{ij}.
\end{equation}

From \eqref{m1} the large $z$-behavior of various shifts can be read out. First $\langle \Phi|g^-]$ and $\langle g^+|\Phi]$ shifts with any $\Phi$  are of order $\mathcal{O}(\frac{1}{z})$, so boundary terms are zero. Next for $\langle g^-|g^+]$ shift, since all higher order terms in $\mathcal{G}^{-+}(z)$ are at most\footnote{All higher order terms have at most one propagator which is $\mathcal{O}(\frac{1}{z})$, while two polarization vectors introduce a $z^2$.} $\mathcal{O}(z)$, we find
\begin{equation}
\begin{aligned}
&\mathcal{G}^{-+}(z)=\mathcal{G}_1^{-+}(z)+\mathcal{O}(z)\\
=&2iz^3k_1\cdot k_{n}gf^{acb}A^{\alpha c}q_{\alpha}+z^2\left(k_1\cdot k_{n}K^{ab}+2gf^{abc}F^{\mu\nu c}k_{1\mu}k_{n\nu}\right)+\mathcal{O}(z),\\
\end{aligned}\label{z-+}
\end{equation}
where
\begin{equation}
K^{ab}=gf^{abc}\{\partial_{\alpha},A^{\alpha c}\}+g^2f^{ace}f^{bde}A_{\alpha}^cA^{\alpha d}.
\end{equation}

If we shift two transverse gluons $\langle g^i|g^j]$
\begin{equation}
\begin{aligned}
\mathcal{G}^{\langle g^i|g^j]}=&\epsilon^i_{\mu} 2izgf^{abc}\eta^{\mu\nu}q_{\alpha}A^{\alpha c}\epsilon^j_{\nu}+\mathcal{O}(1)= 2izg\delta^{ij}f^{abc}q_{\alpha}A^{\alpha c}+\mathcal{O}(1)\\
\end{aligned}
\end{equation}
If $i=j$, $\mathcal{G}^{\langle g^i|g^j]}=\mathcal{O}(z)$, otherwise  $\mathcal{G}^{\langle g^i|g^j]}=\mathcal{O}(1)$.

All the results above are consistent with those of \cite{ArkaniHamed:2008yf}.

\acknowledgments

We thank Rijun Huang and Mingxing Luo for helpful discussions and Rijun Huang for comments on the draft. This work is supported by
Qiu-Shi funding and Chinese NSF funding under contracts No.11031005,
No.11135006, No.11125523 and No.10875103, and National Basic
Research Program of China (2010CB833000).


\bibliographystyle{JHEP}
\bibliography{/Users/jin/Documents/tex/PSUThesis/Biblio-Database}{}

\end{document}